# Ultrasensitive surface plasmon resonance-based biosensor for efficient detection of SARS-CoV-2 Virus in the near-infrared region

R. Runthala[1], N. Grover[2, *] and P. Arora[1, **]

[1]Smart Materials and Sensing Devices Laboratory, Department of Electrical and Electronics Engineering, Birla Institute of Technology and Science Pilani, Rajasthan, 333031, India.
[2]Department of Chemistry, Birla Institute of Technology and Science Pilani, Rajasthan, 333031, India.
*nitika.grover@pilani.bits-pilani.ac.in
**pankaj.arora@pilani.bits-pilani.ac.in

**ABSTRACT:**

This work presents a high-performance, multilayered surface plasmon resonance (SPR)-based sensor designed to enhance performance parameters in the near-infrared (NIR) region through angular interrogation. The multi-layered sensor consists of a bimetallic layer (Aluminum (Al) & Gold (Au)), a dielectric layer ($MgF_2$), and an optimized number of 2D nanomaterial ($MoS_2$) layers. The proposed SPR sensor is numerically designed and analyzed through simulation using the transfer matrix and finite element methods to achieve high sensitivity, figure of merit (FOM), and detection accuracy. The simulated results revealed that the proposed plasmonic sensor (Glass prism/Al/Au/$MgF_2$/$MoS_2$/sensing sample) exhibits a maximum sensitivity of 372°/RIU, FOM of 1690.90 $RIU^{-1}$, and detection accuracy of 4.54 $degree^{-1}$. The outcomes from this work indicate that the developed SPR sensor exhibits excellent detection capabilities for refractive index changes, including those caused by the novel coronavirus, making it a promising prospect for biosensing applications.

 INTRODUCTION

The COVID-19 pandemic has motivated scientists to delve deeper into the sensing area and create cutting-edge biosensors that have the potential to enhance sensing beyond the current state of the art in terms of cost, speed, and accuracy [1]. Several sensors for virus detection have been developed using chemical and electrochemical approaches as their foundations [2][3]. Nonetheless, achieving high performance, accurate recognition, and rapid real-time analysis is a challenging task. The most common diagnostic tests for COVID-19 are RT-PCR, rapid antigen testing, and antibody testing [3][4]. These widely used diagnostic procedures are either costly or difficult to use, and many have low detection accuracy. Optical biosensors, particularly those based on surface plasmon resonance (SPR), overcome most of these drawbacks and provide non-invasive, rapid, label-free, and accurate sensing [5]. Surface plasmons (SPs) are free electron oscillations generated at the metal-dielectric interface due to the phase-matching condition [6]. A coupling prism is usually utilized to match the momentum of the SPs with the momentum of the input p-polarized light. These SPs indicate changes in the analyte's refractive index based on their interaction with the surface of the detecting medium [7].

In a conventional prism-based Kretschmann configuration, a high-refractive-index glass prism is stacked with a plasmonic metal, and a sensing analyte is placed on top of the metal [8]. Gold (Au) and silver (Ag) are the most utilized metal films to excite the SPs on the prism's surface [9], [10]. Au is chemically stable and inert; however, it is an expensive plasmonic metal. Au generates broad SPR curves, which leads to low detection accuracy. Ag provides more sensitivity and narrow curves due to their low optical losses, but it has an oxidation problem [11]. Aluminum (Al) is a plasmonic metal that is not as popular as Au and Ag; still, it has started gaining significant attention due to its availability, compatibility with optoelectronic components, and cost-effectiveness. Moreover, it offers narrower line widths and high sensitivity at higher wavelengths [12]. Recent advancements in SPR-based sensors have led to the use of more optimized heterostructures, such as bimetallic, dielectric-metal-dielectric, and metal-dielectric-metal, which offer better performance than conventional SPR structures [13], [14], [15]. The use of different dielectric materials exhibiting high refractive index, including silicon, barium titanate (BTO), bismuth ferrite (BFO), and indium phosphide (InP), has led to the proliferation of numerous label-free applications like general health monitoring blood glucose, urine, pathogens detection like dengue, malaria, chikungunya, and cancer [16]- [24]. In all these applications, SPR offers its inherent advantages, including ease of use and accurate label-free detection.

Incorporating 2D nanomaterial layers as a bio-recognition element (BRE) on top of the metal-dielectric interface amplifies weak signals from biological events due to their high surface-area-to-volume ratio. This enables us to detect the analyte even at a low concentration [25]. To name a few, black phosphorus (BP), graphene, Mxene, antimonene, $MoS_2$, and fluorinated graphene (FG) as a BRE improved sensitivity and detection accuracy to the next level [26], [27].

Since the pandemic, several reports have described the use of SPR sensors to detect COVID-19. Arun et al. proposed an SPR sensor for detecting SARS-CoV-2 using Ag as plasmonic metal, a single layer of FG, and five layers of carbon nanotubes, achieving a sensitivity of 400°/RIU at 633nm [28]. A combination of dielectric (BFO) and 2D nanomaterial (graphene) is also used to detect COVID-19 using the SPR sensor, giving a sensitivity of 293 °/RIU [29]. Vasimalla et al. numerically investigated the SPR sensor in which dielectric (BTO) is sandwiched between 2D nanomaterials, Franckeite and BP, which offers an FOM of 119.69 $RIU^{-1}$ and sensitivity of 331.54 °/RIU towards COVID-19 [30]. Jacob et al. used a graphene metasurface to efficiently detect COVID-19, claiming a maximum sensitivity of 600 GHz/RIU under wavelength interrogation [31].

Since most contemporary research on the numerical analysis of SPR sensing has focused on improving only one parameter at a time, either sensitivity, detection accuracy, or figure of merit, and has not addressed the simultaneous improvement of multiple parameters, this motivates us to tackle this challenge. Analyzing and comparing the sensor's performance based solely on one parameter provides an incomplete picture, and sometimes even misleads the evaluation process and comparative analysis of sensors. Out of all performance parameters, FOM encompasses a comprehensive set of information, including sensitivity and full width at half maximum (FWHM). Therefore, comparing the sensing performance of SPR sensors based on FOM helps overcome the existing gaps and shortcomings found in the SPR literature. To this end, we envision employing the engineering of a multi-layered plasmonic heterostructure

(Bimetallic layer-dielectric layer-2D nanomaterial layer) in the NIR region with improved sensitivity and FOM simultaneously to a great extent.

The proposed sensor's performance is investigated numerically using the transfer matrix method (TMM) and the finite-element method (FEM) approaches. TMM is a popular method for analyzing the reflection and transmission characteristics of multilayered devices, as it requires no approximations and provides good accuracy simultaneously. Numerical analysis based on an angle interrogation scheme at the telecommunication wavelength (1550 nm) is done using TMM, and the electrical field distribution is analyzed using FEM-based COMSOL Multiphysics. In COMSOL Multiphysics, a TM-polarized plane wave is used under a total internal reflection scheme to excite the SPs at the resonance angle. While continuity is enforced across prism/metal/dielectric/2D material/sensing-medium interfaces for correct field coupling, the outer domain boundaries are assigned scattering boundary conditions to absorb outgoing waves during the angular sweep. Finally, the intensity of reflected light is monitored, verifying the energy balance (incident = reflected + transmitted + absorbed). The proposed sensor can achieve an exceptionally high value of FOM for COVID-19 detection.

## 1. SENSOR'S MODELLING AND PERFORMANCE PARAMETERS

### 1.1 Mathematical Modeling

For an n-layer stacked structure of different materials, a p-polarized light will be illuminated from one side of the device. Several prominent methods are employed to determine its transmission and reflection coefficients, including the transfer matrix method (TMM), field tracing, and resultant wave methods. TMM is used more frequently since it doesn't require approximations, making this method more precise and reliable than other methods.

In this method, all n-layers are stacked along the z-axis. The relation between the tangential fields at the last boundary ($z=z_{n-1}$) and the first boundary ($z=0$) is defined by the characteristic matrix ($M$), where $F_1$ and $G_1$ are the electric and magnetic fields for the first boundary, $F_{n-1}$ and $G_{n-1}$ are the electric and magnetic fields at the last boundary, respectively [32].

$$\begin{bmatrix} F_1 \\ G_1 \end{bmatrix} = M \begin{bmatrix} F_{n-1} \\ G_{n-1} \end{bmatrix} \qquad (1)$$

Moreover, in a multi-layer configuration, the characteristic matrix for the combined structure is defined as

$$M_{ij} = \left(\prod_{k=2}^{N-1} M_k\right)_{ij} = \begin{pmatrix} M_{11} & M_{12} \\ M_{21} & M_{22} \end{pmatrix} \qquad (2)$$

Where

$$M_k = \begin{bmatrix} \cos\beta_k & -\dfrac{i\sin\beta_k}{q_k} \\ -iq_k \sin\beta_k & \cos\beta_k \end{bmatrix} \qquad (3)$$

$$q_k = \dfrac{(\epsilon_k - n_1^2 (\sin\theta_1)^2)^{1/2}}{\epsilon_k} \qquad (4)$$

$$\beta_k = \dfrac{2\pi d_k}{\lambda}\left((\epsilon_k - n_1^2(\sin\theta_1)^2)^{1/2}\right) \qquad (5)$$

Here, $\beta_k$ is the phase constant, and $\theta_1$ is the angle of the incident light. Additionally, $d_k$ is the thickness, and $\epsilon_k$ is the dielectric constant of the $k$th layer [32].

The reflection intensity of p-polarized light for an n-layer model is defined as:

$$R_p = r_p r_p^* = |r_p|^2 \qquad (6)$$

where $r_p^*$ denotes the complex conjugate of the reflection coefficient ($r_p$) for p-polarized light, and $r_p$ is defined as:

$$r_p = \frac{(M_{11}+M_{12}q_n)q_1 - (M_{21}+M_{22}q_n)}{(M_{11}+M_{12}q_n)q_1 + (M_{21}+M_{22}q_n)} \qquad (7)$$

### 1.2 Performance Parameters

The performance of the SPR sensor can be measured with various parameters. Sensitivity, detection accuracy (*DA*), and *FOM* are the most prominent ones; other factors, such as the intensity of reflected light and *FWHM* of the SPR curves, can also be used as decisive parameters. The intensity of the reflected light should be minimal, as denoted by $R_{min}$. Similarly, *FWHM* is expected to be as low as possible. One of the most critical parameters of any sensor is its sensitivity (*S*), which indicates how accurately it can detect minimal changes in the sensing medium. SPR sensor responds to a change in the refractive index of the sensing medium/analyte by showing a shift in the respective resonance angle/wavelength position [6]. A larger shift for a small change in the refractive index results in better sensitivity. Sensitivity for the case of angle interrogation is defined as a ratio of change in resonance angle to change in the refractive index of the respective analyte, which can be mathematically written as,

$$S = \frac{\Delta \theta_{res}}{\Delta n_a} (°/RIU) \qquad (8)$$

where $\Delta \theta_{res}$ = change in resonance angle, and $\Delta n_a$ = change in analyte's refractive index.

The other performance factor is *FWHM*, which is the change or difference in resonance angles at 50% reflected intensity, i.e.

$$FWHM = \theta_2 - \theta_1 = \Delta\theta_{0.5} \qquad (9)$$

Where $\theta_2$ and $\theta_1$ are the resonant angles at 50% intensity, $\Delta\theta_{0.5}$ is the reflectance curve's spectral width at 50% reflectivity, which is nothing but *FWHM*.

Detection accuracy (*DA*) is expressed as the inverse of the *FWHM* value [33].

$$DA = \frac{1}{FWHM} (degree^{-1}) \qquad (10)$$

The *FOM* parameter provides a clear and intuitive understanding of the SPR sensor's performance. It is one of the most critical parameters, defined as a ratio of *S* and *FWHM* [33].

$$FOM = \frac{S}{FWHM} (RIU^{-1}) \qquad (11)$$

### 2. RESULTS AND DISCUSSION

Our proposed Kretschmann configuration-based SPR sensor is a four-layer heterostructure, consisting of a glass prism, a bimetallic layer of Al and Au, a dielectric layer (MgF2), and finally, a 2D nanomaterial (MoS2) as the BRE layer. Figure 1 shows the proposed schematic and required arrangement for SPR sensing. In this setup, a monochromatic light source at 1550 nm is incident on the glass prism through a polarizer, and the layered architecture is placed on a rotating base, allowing the incident angle to be varied, thereby enabling the angle interrogation technique. The reflected light is detected by a photodetector. The photodetector here refers to an idealized detector element included in the simulation to monitor the reflected optical power. From an experimental perspective, the detector can be considered conceptually equivalent to an InGaAs photodiode, which is commonly used for detection at around 1550 nm due to its high responsivity and low noise. However, within our simulation framework, the detector response was treated as ideal, with unit quantum efficiency and flat spectral responsivity, measuring only the optical intensity.

*Prism:*

A metal-coated glass prism is used in this basic Kretschmann configuration. The function of the glass prism is to increase the momentum of incoming *p*-polarized light. Here, we have used sodium fluoride (NaF) as a glass prism with the lowest refractive index value. Moreover, $R_{min}$ values obtained in reflection characteristics using NaF prism are minimal compared to other commonly used glass prisms, e.g., calcium fluoride ($CaF_2$), borosilicate crown glass (BK-7), and barium fluoride ($BaF_2$) [34].

*Plasmonic Metal:*

A plasmonic metal layer is placed on the top of the glass prism. Selecting a plasmonic metal with its optimal thickness is crucial for achieving high sensing parameters. Ag is commonly used as a plasmonic metal for optical applications due to its lack of interband transitions [11]. Still, Ag has a significant disadvantage since it is prone to oxidation. In a similar vein, Au is a commonly used plasmonic material in various applications because of its intrinsic features, such as chemical inertness and biocompatibility [35]. As a result, Au has been widely used as the primary plasmonic material in various theoretical and experimental studies [10], [36]. Copper (Cu) is a cost-effective alternative to Au and Ag; however, it produces fewer surface plasmons (SPs) compared to Au and Ag [37]. Recently, Al has gained significant attention as a plasmonic metal. Al has become a superior choice due to its availability and low cost [33].

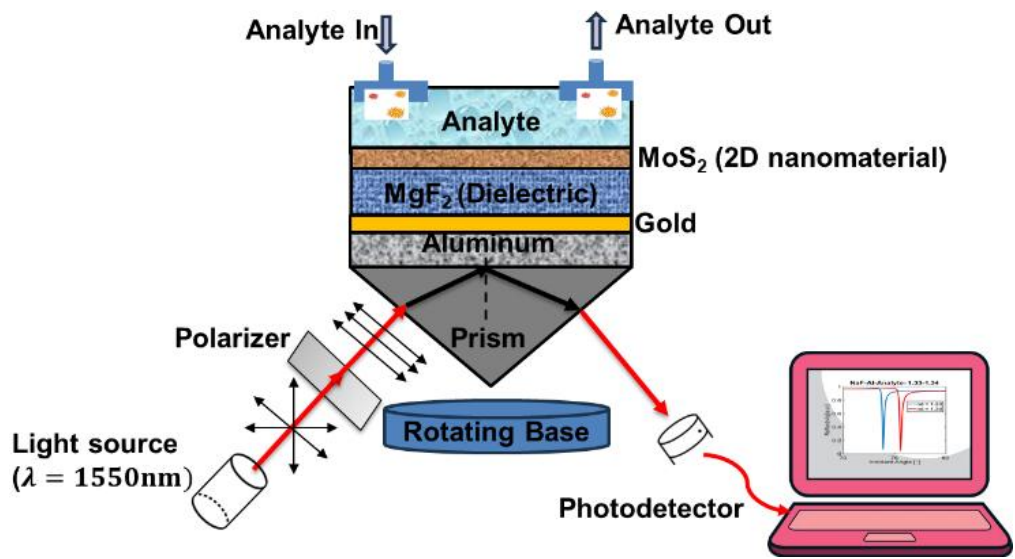

Fig. 1. Schematic of the proposed multi-layered SPR-based sensor

To select the most suitable plasmonic metal for our targeted sensor, we choose *FWHM* and $R_{min}$ as decisive parameters. Ideally, the values of these parameters should be as low as possible. Figure

2(a) shows the reflection characteristics for different metals with the optimized thickness (Al=30 nm, Au=45 nm, Ag=45 nm, and Cu=50nm) in a conventional Kretschmann configuration. As evident from Figure 2(a) and its inset view in Figure 2(b), Aluminium (Al) exhibits the lowest reflectance minimum ($R_{min}$ = 0.001763) and the narrowest resonance dip with a minimum FWHM of 0.21, compared to Ag, Au, and Cu, which show $R_{min}$ values of 0.006465, 0.02937, and 0.03893, and corresponding *FWHM* values of 0.32, 0.49, and 0.33, respectively. As a result, Al with a thickness of 30 nm is taken for further simulations. Additionally, a very thin layer of Au (5 nm) is considered on top of Al to prevent oxidation [7]. Refractive indices of these plasmonic metals are calculated using Drude-Lorentz's model with plasma wavelength ($\lambda_p$) and collision wavelength ($\lambda_c$) mentioned in [38], [39].

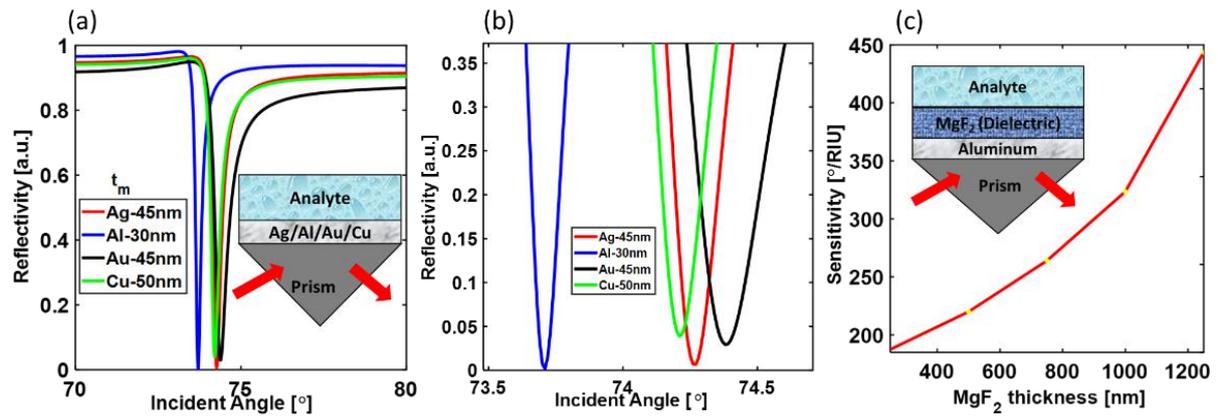

Fig. 2. (a) Reflection characteristics for a conventional Kretschmann's configuration with different plasmonic metals (Ag, Al, Au, and Cu); (b) Inset view of Fig. 2(a) for reflection characteristics of different metals; (c) Variation in the Sensitivity value with thickness variations of $MgF_2$ layers

*Dielectric:*
Many dielectric materials, e.g., BTO, titanium dioxide ($TiO_2$), platinum diselenide ($PtSe_2$), and silicon dioxide ($SiO_2$), have been used recently in SPR-based heterostructures to improve sensitivity [17], [40]. In this work, we use magnesium fluoride ($MgF_2$) as a dielectric material to improve the performance parameters. According to earlier research, $MgF_2$ can be very helpful in SPR-based sensors operating in the NIR region, as it is immune to radiation damage at the NIR wavelengths. Moreover, $MgF_2$ has a minimal thermo-optic coefficient, which helps to maintain consistent sensor performance against temperature variations [41]. The structure is initially analyzed using an analyte with refractive indices of 1.33 and 1.34, respectively. The thickness of $MgF_2$ is varied from 250 nm to 1250 nm, with increments of 250 nm, to achieve the optimized thickness. There is a gradual rise in sensitivity with an increase in thickness, as shown in Figure 2(c). However, the SPR angle also starts to increase, as shown in Table 1, and approaches 90°, with a value of 87.26° at 1250 nm. The SPR angle closer to 90° is not realizable experimentally [6], so a thickness value just below 1250 nm, i.e., 1000 nm, is chosen, which gives a sensitivity of 322.5° at an SPR angle of 84.58°.

**Table 1. Effect of $MgF_2$ Thickness on the Sensitivity and SPR Angle**

| Thickness [nm] | Sensitivity [°/RIU] | SPR Angles[°] |
|---|---|---|
| 250 | 186.5 | 76.58, 78.44 |
| 500 | 219.5 | 78.49, 80.68 |
| 750 | 263.5 | 79.98, 82.61 |
| 1000 | 322.5 | 81.35, 84.58 |
| 1250 | 440 | 82.86, 87.26 |

*2D Nanomaterial:*

The topmost layer of the SPR sensor is coated with 2D nanomaterials, which facilitate excellent binding with the analyte due to their large surface-to-volume ratio [25]. Out of the wide range of 2D nanomaterials, we use $MoS_2$ as a 2D nanomaterial. $MoS_2$ is frequently employed in SPR sensing due to its unique benefits, including a large band gap (1.8 eV), a wide work function, and a higher light absorption efficiency (5-6%) [42]. Additionally, the large-area growth of $MoS_2$ is possible using chemical vapor deposition, which facilitates the development of $MoS_2$-based sensors. To optimize the number of layers of $MoS_2$, layers are incremented from 1 to 20 layers, and respective performance parameters (*S* and *FOM*) are numerically analyzed, as shown in Figure 3(a), where the thickness of each $MoS_2$ monolayer is 0.65 nm. With the increase in $MoS_2$ layers, sensitivity rises linearly, but FOM values decrease sharply. A good trade-off between the S and FOM is found to be at the value of 10 layers. Therefore, the optimized number of layers for $MoS_2$ is chosen to be 10 for the targeted sensor.

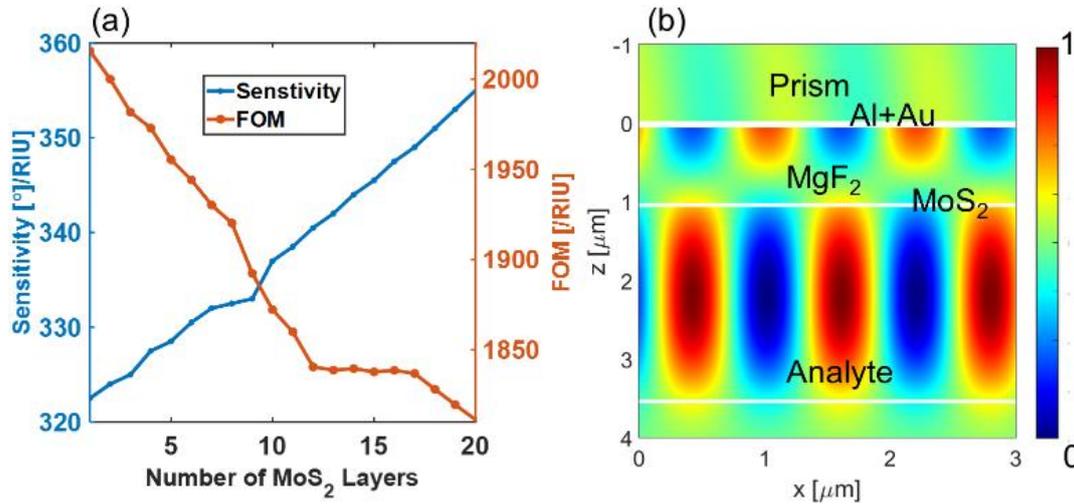

Fig. 3. (a) Effect of the number of $MoS_2$ layers on sensitivity and FOM; (b) Electric field distribution for the proposed sensor at the resonance angle

Figure 3(b) shows the electric field distribution for the engineered SPR sensor at the resonance angle. The field is found to propagate through the analyte due to a first-order resonance mode, in addition to the fundamental SP mode at the metal interface, resulting in enhanced interaction and improved sensing parameters.

**Table 2. Design parameters considered for the proposed sensor**

| Layer | Thickness[nm] | RI |
|---|---|---|
| Prism | - | 1.3194 |
| Al (metal) | 30 | 0.4598 -14.4882i |
| Au (metal) | 5 | 0.57470-9.6643i |
| $MgF_2$ (dielectric) | 1000 | 1.3705 |
| $MoS_2$ (2D nanomaterial) | 10*0.65 | 3.647 |

Table 2 summarizes the optimized design parameters for the respective layers, including RI and thickness of all the stacked layers.

After selecting each layer with its optimized thicknesses, the possible fabrication steps for the proposed multilayer configuration are shown in Figure 4. A 30 nm Al metal layer and 5 nm Au will be deposited on a clean glass prism using a thermal evaporator. A 1,000 nm layer of dielectric material (MgF$_2$) will be deposited on the metal-coated glass prism using a sputtering technique. In the end, 10 layers of 2D nanomaterial (MoS$_2$) will be deposited using the chemical vapor deposition method.

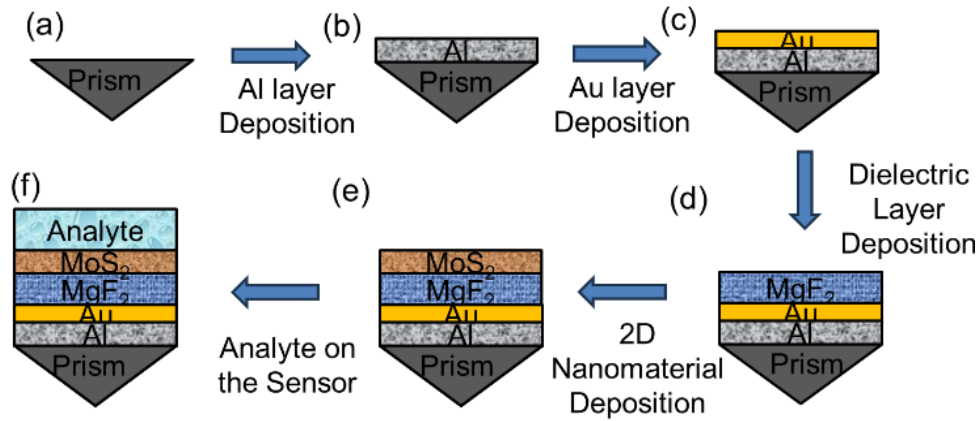

Fig. 4. Possible fabrication steps for the proposed SPR sensor (schematic not to scale)

Selecting the best material for an individual layer and figuring out its ideal thickness was the first step in designing an SPR sensor. To verify the importance of each layer proposed in the engineered sensor, the following four different configurations are employed for sensing applications:

Configuration I: Prism-Al(30nm)-Analyte
Configuration II: Prism-Al(30nm)-Au(5nm)-Analyte
Configuration III: Prism-Al(30nm)-Au(5nm)-MoS$_2$(10layer)-Analyte
Configuration IV: Prism-Al(30nm)-Au(5nm)- MgF$_2$(1000nm)-MoS$_2$(10layer)-Analyte

*FOM, R$_{min}$*, and *S* are used to compare the performance of these configurations, which are presented in Table 3.

**Table 3. Performance Analysis of Proposed Configurations for SPR-Based Sensor**

| Configuration | I | II | III | IV |
|---|---|---|---|---|
| S [°/RIU] | 168 | 170 | 176 | 337 |
| FWHM [°] | 0.21 | 0.205 | 0.21 | 0.18 |
| FOM [RIU$^{-1}$] | 800 | 829.26 | 838.09 | 1872.22 |
| R$_{min}$ [a.u.] | 0.0017 | 0.0671 | 0.0611 | 0.001 |
| SPR Angles [°] | 73.71, 75.39 | 73.88, 75.58 | 74.37, 76.13 | 81.60, 84.97 |

Configuration I offers a sensitivity of 168°/RIU and an FOM of 800 RIU$^{-1}$ (Figure 5(a)). For configuration II, adding a 5 nm Au layer to protect Al from oxidation slightly enhances the sensitivity to 170°/RIU and increases the FOM to 829 RIU$^{-1}$ (Figure 5(b)). However, this improvement leads to an undesired increase in the value of R$_{min}$. Configuration III incorporates a 2D nanomaterial (MoS$_2$) layer, which acts as a binding layer, resulting in further improving the sensitivity (176°/RIU) due to the increased number of binding sites for the analyte (Figure 5(c)). Finally, in configuration IV, the addition of an MgF$_2$ layer, which acts as a dielectric, is inserted between the bimetallic and binding layers, leading to a significant improvement in performance parameters, with S reaching 337°/RIU and FOM of 1872 RIU$^{-1}$ (Figure 5(d)). Therefore, configuration IV is found to be most suitable for demonstrating biosensing applications.

Samples with different COVID-19 concentrations are considered for testing the sensing capabilities of the engineered SPR-based sensor (configuration IV) [28]. A thin layer of PBS solution is considered below the COVID sample to activate the sensing surface. The refractive indices of samples with different COVID-19 concentrations in the NIR region are calculated using the equation below:

$$n_{analyte} = n_{PBS} + C_{analyte}(\delta n/\delta c) \qquad (12)$$

where $c_{analyte}$ is the concentration of the COVID-19 sample, $n_{PBS}$ is the refractive index for the used PBS solution of value 1.3348, and $\delta n/\delta c = 0.181 cm^3/g$ is the rate of change of refractive index modification with concentration.

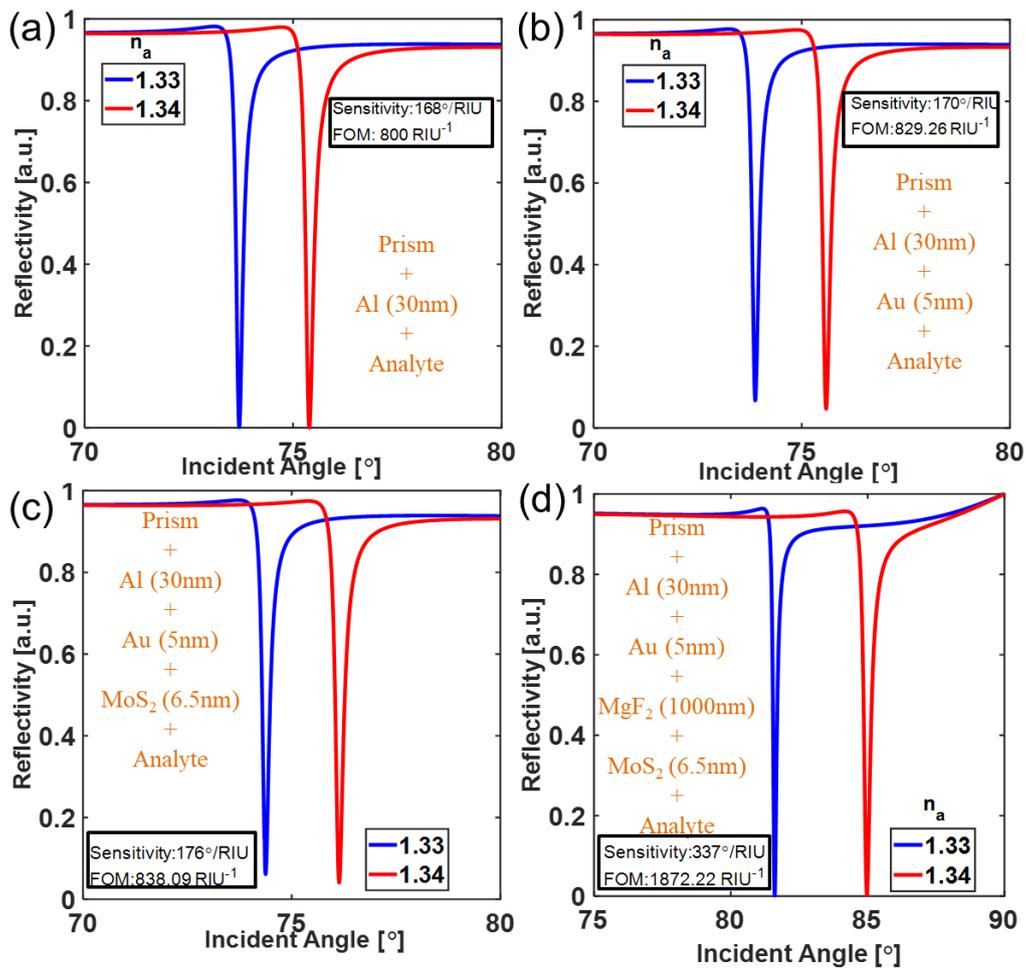

Fig. 5. Comparison of the performance parameters for proposed structures

Since the refractive index of these infected samples falls within our tested analyte range of 1.33 to 1.34, the proposed sensor is a potential candidate for efficient performance in the aforementioned application. For real-time applications, biomolecular recognition components, such as antibodies, aptamers, or molecularly imprinted polymers specific to SARS-CoV-2 antigens, must be functionalized on the $MoS_2$ surface to achieve target-specific binding. Functionalization can be achieved through non-covalent interactions for aptamer attachment or through well-known chemical linkers (such as EDC/NHS chemistry for antibody immobilization). This biorecognition layer would seamlessly integrate on top of the $MoS_2$ surface without significantly disrupting the plasmonic behavior. This bio-interface should be included to transform the sensor from a generic refractive index detector into a highly specific diagnostic platform. Furthermore, SPR-based sensors are highly sensitive to temperature fluctuations, humidity, and biofouling. Encapsulating the sensor in an environmental chamber with controlled temperature will shield it from external fluctuations, thereby maintaining a stable environment around the sensor.

Figure 6 (a) shows the reflection characteristics of the sensor with variations in different concentrations. An increase in the SPR angle is observed with increasing concentration, as depicted in Figure 6(b). A sensitivity value of 320 °/RIU for a concentration value of 1.953 nM and a sensitivity value of 399.2 °/RIU for a concentration of 62.5 nM are observed in Figure 6(a), respectively. At the same time, a FOM of 1690.90 $RIU^{-1}$ is achieved in sensing a COVID-19 sample of 62.5 nm concentration from a healthy sample with a sensitivity of 372°/RIU. Table 4 presents the performance parameter analysis for samples of varying concentrations.

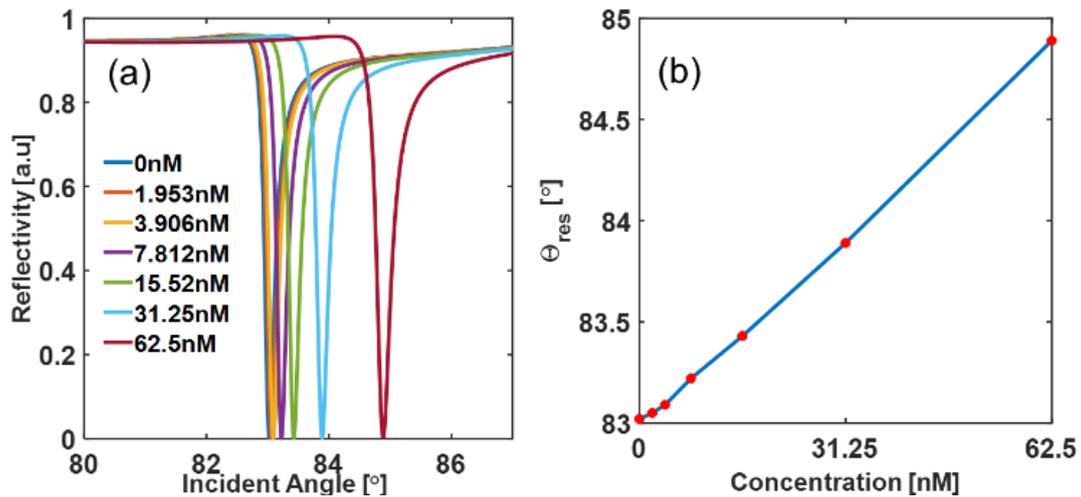

Fig. 6: (a) Reflection characteristics of the proposed sensor with the variation in COVID-19 sample concentrations, (b) Plot between the SP resonance angle and the COVID-19 sample concentrations.

**Table 4. Performance Parameters Analysis for Different COVID-19 Sample Concentrations**

| Concentration [nM] | Respective RI values | SPR Angle [°] | Sensitivity [°/RIU] | DA [degree$^{-1}$] | FOM [RIU$^{-1}$] |
|---|---|---|---|---|---|
| 0 | 1.3348 | 83.025 | - | 4.76 | - |
| 1.953 | 1.3371 | 83.057 | 320 | 4.73 | 1516.58 |
| 3.906 | 1.3417 | 83.09 | 330 | 4.69 | 1549.29 |
| 7.812 | 1.3417 | 83.22 | 325 | 4.60 | 1497.69 |
| 15.52 | 1.3441 | 83.425 | 341.66 | 4.44 | 1518.51 |
| 31.25 | 1.3465 | 83.887 | 355.38 | 4.11 | 1462.48 |
| 62.5 | 1.3491 | 84.885 | 399.2 | 3.42 | 1367.12 |

The initial phase of sensor design involves thoroughly examining mathematical models to define optimal performance criteria and determine the precise physical requirements for manufacturing. However, despite these meticulous preparations, the actual performance of the sensor often deviates from its theoretical expectations. This difference can be attributed to various factors, including fabrication complexities and the packaging techniques used for the sensor. Each of these factors significantly impacts the sensor's ultimate operational characteristics.

In sensor development, variations in thickness often occur unexpectedly during the fabrication of thin layers. Additionally, variability in biological samples (such as COVID concentrations) and material properties (refractive indices) can also occur during real-time measurements. Moreover, temperature can also influence sensitivity through changes in the refractive index ($dn/dT$), metal permittivity, and dielectric layer thickness due to thermal expansion. Mitigating these deviations and their potential effects on sensor performance requires a comprehensive investigation. To investigate this, we perform precise analytical computations to assess the implications of material thickness, refractive indices, and concentration of biological samples methodically. Table 5 presents the performance parameters when material thicknesses are varied by ±1%, ±2%, ±3%, and ±4%. Notably, the results illustrate that sensor performance parameters exhibit a minimal change in response to these thickness variations. This consistency indicates that the sensor design effectively mitigates the impact of fabrication errors arising from unintended thickness variations.

**Table 5. Deviation of Parameters from the Expected Values with Thickness Variations**

| % Variation in thickness | Sensitivity [°/RIU] | FWHM [°] | FOM [RIU$^{-1}$] |
|---|---|---|---|
| +4% | 373 | 0.220 | 1695.45 |
| +3% | 373 | 0.225 | 1657.77 |
| +2% | 373 | 0.220 | 1695.45 |
| +1% | 372 | 0.220 | 1690.90 |
| Ideal/Expected | 372 | 0.220 | 1690.90 |
| -1% | 372 | 0.220 | 1690.90 |
| -2% | 372 | 0.220 | 1690.90 |
| -3% | 371.5 | 0.225 | 1651.11 |
| -4% | 371 | 0.220 | 1686.36 |

Table 6 presents the performance parameters when material properties (refractive indices) are varied within a range of ±0.25% to ±1.0%. The sensor's sensitivity remains largely unaffected across the entire range of deviations. A minor variation is observed in the FWHM, which in turn slightly impacts the FOM.

**Table 6. Effect of Refractive Index (RI) Variation for Different Layers used in the Proposed Device**

| % change in RI | Sensitivity [°/RIU] | FWHM [°] | FOM [RIU$^{-1}$] |
|---|---|---|---|
| +1.0% | 372 | 0.215 | 1730.23 |
| +0.75% | 372 | 0.22 | 1690.90 |
| +0.50% | 371 | 0.22 | 1686.36 |
| +0.25% | 372 | 0.225 | 1653.33 |
| *Ideal/Expected* | *372* | *0.220* | *1690.90* |
| -0.25% | 372 | 0.225 | 1653.33 |
| -0.50% | 373 | 0.225 | 1657.77 |
| -0.75% | 373 | 0.23 | 1621.73 |
| -1.0% | 372 | 0.23 | 1617.39 |

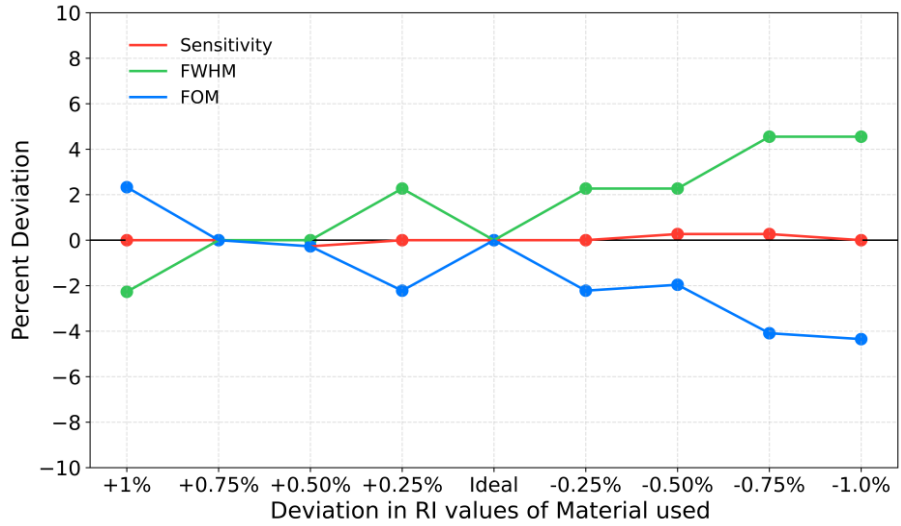

Fig. 7: Graphical summary representing the percent deviation in performance parameters when the RI of materials is varied

However, these deviations become significant only at the extreme ends of the variation spectrum (±1.0%), which are not only challenging to achieve under standard laboratory conditions but also represent rare occurrences in practical scenarios. These findings underscore the proposed sensor's robustness and reliability in the face of realistic fluctuations in material properties. Figure 7 graphically shows these deviations; the sensitivity (red line) shows almost no deviations, while the FWHM (green line) increases with deviations, resulting in a proportional decrease in FOM (blue line) as well.

Table 7 presents the effect on sensing parameters due to deviations in the RI values of the biological sample under consideration. At the same time, a graphical summary of this analysis is provided in Figure 8. Sensitivity exhibits an almost linear change; FWHM deteriorates with variations in RI, resulting in a corresponding effect on FOM. The analysis reveals that the proposed sensor demonstrates a high degree of robustness against such variations in the RI of the sensing medium, maintaining nearly consistent performance across the entire deviation range. This resilience ensures reliable and accurate sensor functionality in real-world applications.

**Table 7. Effect of Refractive Index Variation in Biological Sample**

| Deviation in RI | Sensitivity [°/RIU] | FWHM [°] | FOM [RIU$^{-1}$] |
|---|---|---|---|
| +0.0003 | 379 | 0.230 | 1647.82 |
| +0.0002 | 377 | 0.225 | 1675.55 |
| +0.0001 | 375 | 0.225 | 1666.66 |
| *Ideal /Expected* | *372* | *0.220* | *1690.90* |
| -0.0001 | 370 | 0.225 | 1644.44 |
| -0.0002 | 368 | 0.220 | 1672.72 |
| -0.0003 | 365 | 0.220 | 1659.09 |

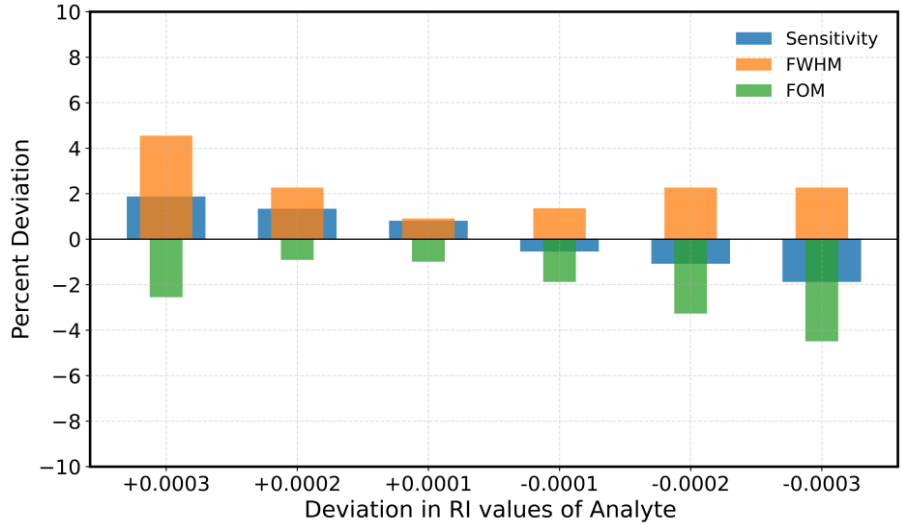

Fig. 8: Graphical summary representing the percent deviation in performance parameters when the RI of the analyte is varied

Table 8 presents a comparative analysis of state-of-the-art and current work, clearly demonstrating that the present work on SPR sensing for COVID-19 is far ahead of its contemporary counterparts. In a bimetallic configuration (Cu+Ni) with $TiO_2$ as a dielectric and BP as a 2D nanomaterial for COVID-19 samples sensing, an excellent sensitivity of 502°/RIU but at the cost of a poor FOM of only 100 $RIU^{-1}$, as shown in Table 8 [40]. Similarly, a four-layer structure consisting of Ag, CNT, FG, and a Thiol group offered a sensitivity of 390°/RIU but a poor value of FOM of 87.95 $RIU^{-1}$ [43]. A poor value of FOM was obtained in other works, too, as mentioned in Table 8 [30], [44], [45], [46], [47], [48]. The proposed work offers an FOM of 1690 $RIU^{-1}$ with a comparable sensitivity value.

**Table 8. Performance Comparison of the Proposed SPR Sensor with Recent Works on SARS-CoV-2 Detection**

| Structure | Sensitivity (°/RIU) | FOM ($RIU^{-1}$) | References |
|---|---|---|---|
| Al-Au-$MgF_2$-$MoS_2$ | 372 | 1690.90 | Current Work |
| Ag-Fr-BTO-BP | 331 | 119.69 | 2024[30] |
| Cu-Ni-$TiO_2$-BP | 502 | 100.56 | 2024[40] |
| $TiO_2$-Ag-BP-Gr | 390 | 87.95 | 2023[43] |
| Ag-CNT-FG-Thiol | 400 | 76 | 2023[44] |
| Ag-$TiO_2$-Mxene | 346 | 119 | 2023[45] |
| Ag-$S_3N_4$-$MoS_2$ | 375.01 | 38.34 | 2025[46] |
| Ag-$BaTiO_3$-$WS_2$ | 450 | 128.57 | 2025[47] |
| Ag-Ni-$BaTiO_3$ | 533 | 115 | 2025[48] |

## CONCLUSIONS

In this work, we proposed a multilayer SPR sensor with enhanced sensing parameters in the near-infrared region. The engineered sensor consists of a bimetallic layer of Al and Au with an optimized thickness of 30 nm and 5 nm, respectively, $MgF_2$ as a dielectric layer of thickness 1000 nm, and ten layers of $MoS_2$ as a binding layer to give the best performance parameters at 1550 nm wavelength. The proposed plasmonic sensor is set to become a new landmark in achieving a sensitivity of 372°/RIU and an ultra-high value of FOM 1690.90 $RIU^{-1}$ for biosensing applications. To the best of our knowledge, the proposed sensor outperforms current state-of-the-art results published till now in the field of numerical analysis of SPR. Even though the majority of this work is simulation-based, the proposed multilayered SPR biosensor exhibits strong design characteristics that make it feasible to fabricate in practice. The selected materials, Al, Au, $MgF_2$, and $MoS_2$, are readily accessible and are compatible with the popular thin-film deposition methods of sputtering, chemical vapor deposition, and electron-beam evaporation. The integration of 2D nanomaterials (such as $MoS_2$ layers) can be accomplished by methods like liquid-phase deposition or mechanical exfoliation, which have demonstrated promising scalability in recent studies. Utilizing cleanroom or nanofabrication facilities at BITS Pilani could expedite the development of the proposed sensor prototype. Achieving accurate angular interrogation in a small, portable package, maintaining interface quality across layers, and guaranteeing uniformity of ultrathin layers are some of the major hurdles expected in the shift from simulation to real-world deployment. However, we also analyzed our numerically calculated results for fabrication feasibility issues by explicitly incorporating fabrication errors, and the outcomes showed that the proposed sensor's response is almost the same and lacks any absurdity, making this sensor stable against fabrication errors and robust in its performance. The experimental validation of the design and the development of a miniature SPR sensing apparatus suitable for field or clinical settings will be the primary goals of future work. We hope that this comprehensive numerical analysis, which provides high FOM as well as robustness against variations in material properties, will serve as a valuable reference for current and future research groups working in surface plasmonics, particularly in the real-time monitoring of biological samples and chemical analytes in near-infrared sensing platforms. In this work, we have incorporated and analyzed several key parameters within our simulations, while acknowledging that other factors can be further explored experimentally. This approach not only reduces redundant efforts but also provides a foundation upon which subsequent experimental studies can build, thereby accelerating progress in the field.